\documentclass[notitlepage,11pt]{article}

\usepackage{amsfonts}
\usepackage{geometry}
\usepackage[onehalfspacing]{setspace}
\usepackage{graphicx}
\usepackage{amssymb}
\usepackage{amsthm,amsmath}
\usepackage[dvipsnames]{xcolor}
\usepackage {ulem}
\usepackage{enumerate}
\usepackage{enumitem}
\usepackage{dsfont}
\usepackage{natbib}
\usepackage{hyperref}
\usepackage{multirow}
\usepackage{tablefootnote}

\usepackage{tabularx}
\usepackage{multicol}

\usepackage{subcaption}

\hypersetup{colorlinks=true, linkcolor=red,          
    citecolor=blue,        
    filecolor=magenta,      
    urlcolor=cyan           
}

\usepackage[T1]{fontenc}

\usepackage{tgtermes}
\usepackage{cmbright}

\usepackage{bm}

\begin{document}

\title{\begin{center}
\vspace{-3cm} 
    \noindent\fbox{%
        \parbox{\dimexpr\textwidth-35\fboxsep-2\fboxrule\relax}{%
             \small {\fontfamily{ptm}\selectfont Please note that an e-comment is not an \textit{AER} publication and should not be cited as such. This e-comment should be cited as:\\ 
            Demeulemeester, Tom. 2026. ``E-comment on `What's the Matter with Tie-Breaking: Improving Efficiency in School Choice.'{''} 
            \\Posted at \url{https://www.aeaweb.org/articles?id=10.1257/aer.98.3.669}.}
        }%
    }\\[1em]
\end{center}E-comment on `What's the Matter with Tie-Breaking: Improving Efficiency in School Choice'\thanks{I would like to thank Aytek Erdil for his help in identifying this correction. Additionally, I would like to thank Lars Vilhuber for his help in reviewing the updated code. I gratefully acknowledge the financial support from the Swiss National Science Foundation (SNSF) through Project 100018-231777.}}
\author{Tom Demeulemeester\thanks{Department of Quantitative Economics, Maastricht University.  (email: tom.demeulemeester@maastrichtuniversity.nl)}}

\maketitle
\vspace{-1em}
\begin{abstract}
The code that was used in \cite{erdil2008s} to compute stable improvement cycles sometimes generated unstable matchings. I identify the minor bug in their code that caused this issue, and I present a corrected implementation. While the general insights from the computational experiments obtained by \cite{erdil2008s} persist, the true fraction of improving students is slightly smaller than reported, while their average improvement in rank is larger than reported. All theoretical findings in \cite{erdil2008s} are unaffected. 
\end{abstract}


\section{Counterexample} 

\cite{erdil2019replication}'s replication package contains a code to implement the stable improvement cycles algorithm. The following is an example in which a minor bug in their code results in an unstable matching.

Consider an instance with four agents $\{1,2,3,4\}$, four unit-capacity objects $\{a,b,c,d\}$, and the following preferences and priorities.

\[
    \begin{tabular}{cccc}
        $\succ_1$ & $\succ_2$ & $\succ_3$ & $\succ_4$\\ \hline
        c & b & a & a\\
        a & c & b & c\\
        b & a & c & b\\
        d & d & d & d
    \end{tabular}
    \qquad \qquad
    \begin{tabular}{cccc}
        $\succeq_a$ & $\succeq_b$ & $\succeq_c$ & $\succeq_d$\\ \hline
        2,3,4 & 4 & 3 & 4 \\
        1 & 2 & 1,2 & 3\\
         & 3 & 4 & 2\\
         & 1 &  & 1
    \end{tabular}
\]
If the ties are broken according to the ordering $(1,2,3,4)$, then the deferred acceptance mechanism would find the matching
\[
\mu = \begin{pmatrix}
    1&2&3&4\\
    d&a&c&b\\
\end{pmatrix}.
\]

The stable improvement cycles algorithm could identify two cycles: either agents 2 and 4 first exchange $a$ and $b$, or agents 2 and 3 exchange $a$ and $c$. In either case, no further stable improvement cycles can be implemented afterwards, resulting in one of the following two stable matchings:

\[\mu^1 = \begin{pmatrix}
    1&2&3&4\\
    d&b&c&a\\
\end{pmatrix}, \qquad \qquad \mu^2 = \begin{pmatrix}
    1&2&3&4\\
    d&c&a&b\\
\end{pmatrix}.\]

When applying the published code by \cite{erdil2019replication} to matching $\mu$, at first the cycle between agents 2 and 4 is resolved. However, afterwards, it also resolves a cycle between agents 2 and 3, in which they exchange $c$ and $b$, resulting in the following matching:
\[
\mu' = \begin{pmatrix}
    1&2&3&4\\
    d&c&b&a\\
\end{pmatrix}.
\]
Note that this matching is not weakly stable, because $(2,b)$ is a blocking pair. 

\section{Correction code}
The error is caused by inaccurate updating in the code by \cite{erdil2019replication} of which schools the students prefer to their currently matched school. More specifically, imagine a student $i$ being matched to school $x$ before resolving a stable improvement cycle, and to school $y$ that they prefer to $x$ after resolving the cycle. The current implementation of the code will only remove student $i$ from the set $D_y$, which is the subset of the students with the highest priority among all students who prefer $y$ over their currently assigned school. However, the proposals of student $i$ to all schools that they rank higher than $x$, but lower than $y$ are not removed in the current implementation. As such, the current implementation of the code does not exclude that student $i$ can still be involved in cycles for all schools that they rank higher than $x$, but lower than $y$. 

The counterexample illustrates this, as student 2 receives their first-ranked school $b$ after the first stable improvement cycle, and they will be removed from $D_b$. In the next iteration, student 2 will wrongfully belong to $D_c$ in the current implementation of the code. As a result, student 2 will therefore receive their second-ranked school $c$ in the next iteration, which they prefer less than $b$ (but still more than $a$, the school they were initially matched to).

This can be resolved in the current implementation of the code by replacing the following lines in the definition of the function \texttt{improve\_allocations}
\begin{verbatim}
    proposals[school2][rank].remove(student)
    if len(proposals[school2][rank]) == 0:
            proposals[school2].pop(rank)
\end{verbatim}
with the following few lines:
\begin{verbatim}
    pref_index_old_school = N[student].index(school1)
    pref_index_new_school = N[student].index(school2)
    for i in range(pref_index_new_school, pref_index_old_school):
        school3 = N[student][i]
        rank = A[school3][student]
        proposals[school3][rank].remove(student)
        if len(proposals[school3][rank]) == 0:
            proposals[school3].pop(rank)

\end{verbatim}

An updated version of the code is available here: \url{http://doi.org/10.3886/E248175V1}.

\section{Impact on computational results}
We evaluate the effects of the described bug on the computational experiments in \cite{erdil2008s} by running the originally implemented code and the corrected code for 1,000 students, 20 schools, and differing values of $\alpha$ and $\beta$ (averaged over 200 instances). In short, the parameter $\alpha$ captures the correlation in the students' preferences, while the parameter $\beta$ captures how sensitive the students' preferences are to locational proximity. For more details about the data generation process, we refer to \cite{erdil2008s}.

\begin{figure}[!htb]
\centering
\includegraphics[width=.95\linewidth]{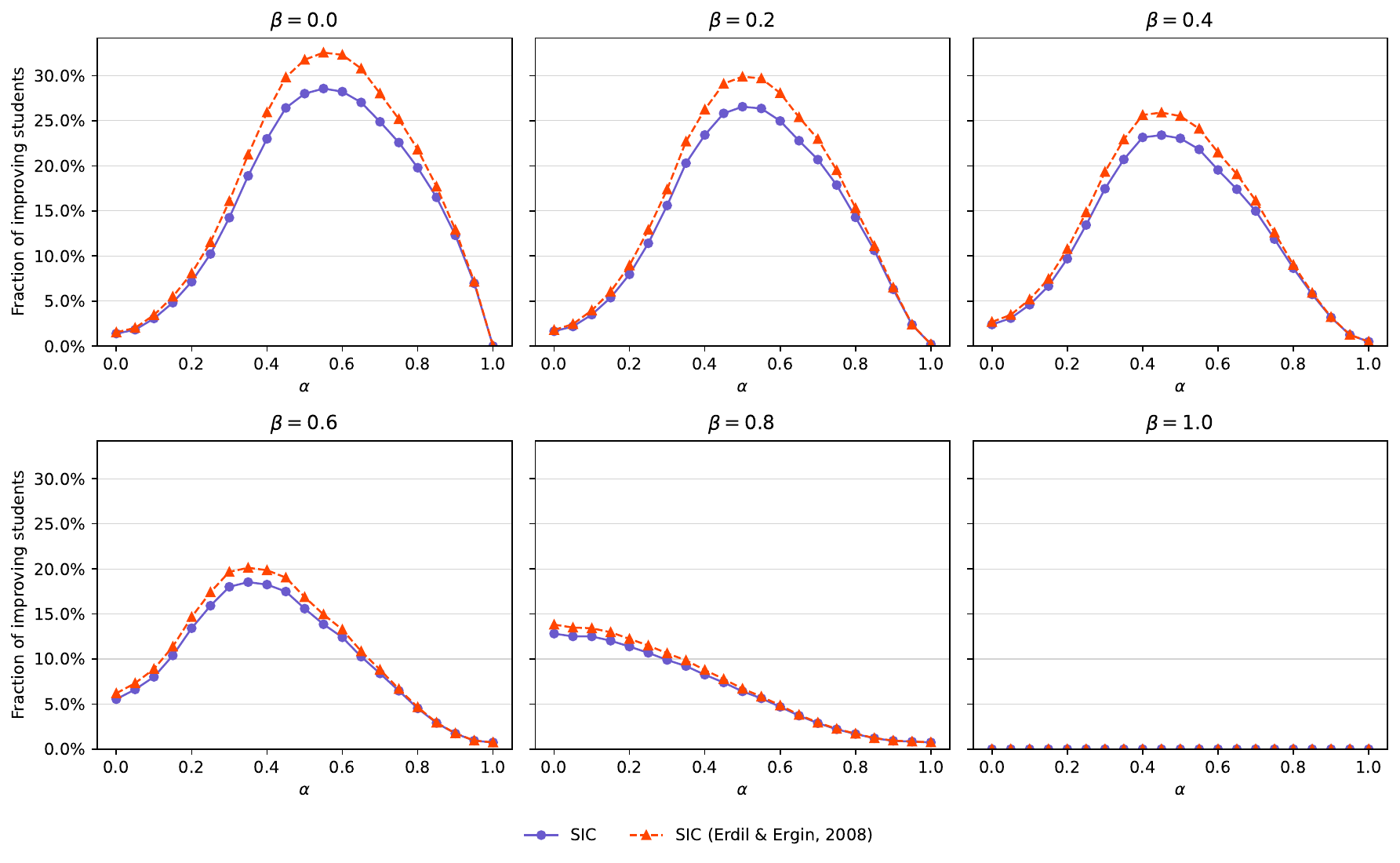}
\caption{Percent of improving students as a function of $\alpha$ (1,000 students and 20 schools).}
\label{fig:fraction_impr_DA}

\end{figure}

\begin{figure}[!htb]
\centering
\includegraphics[width=.95\linewidth]{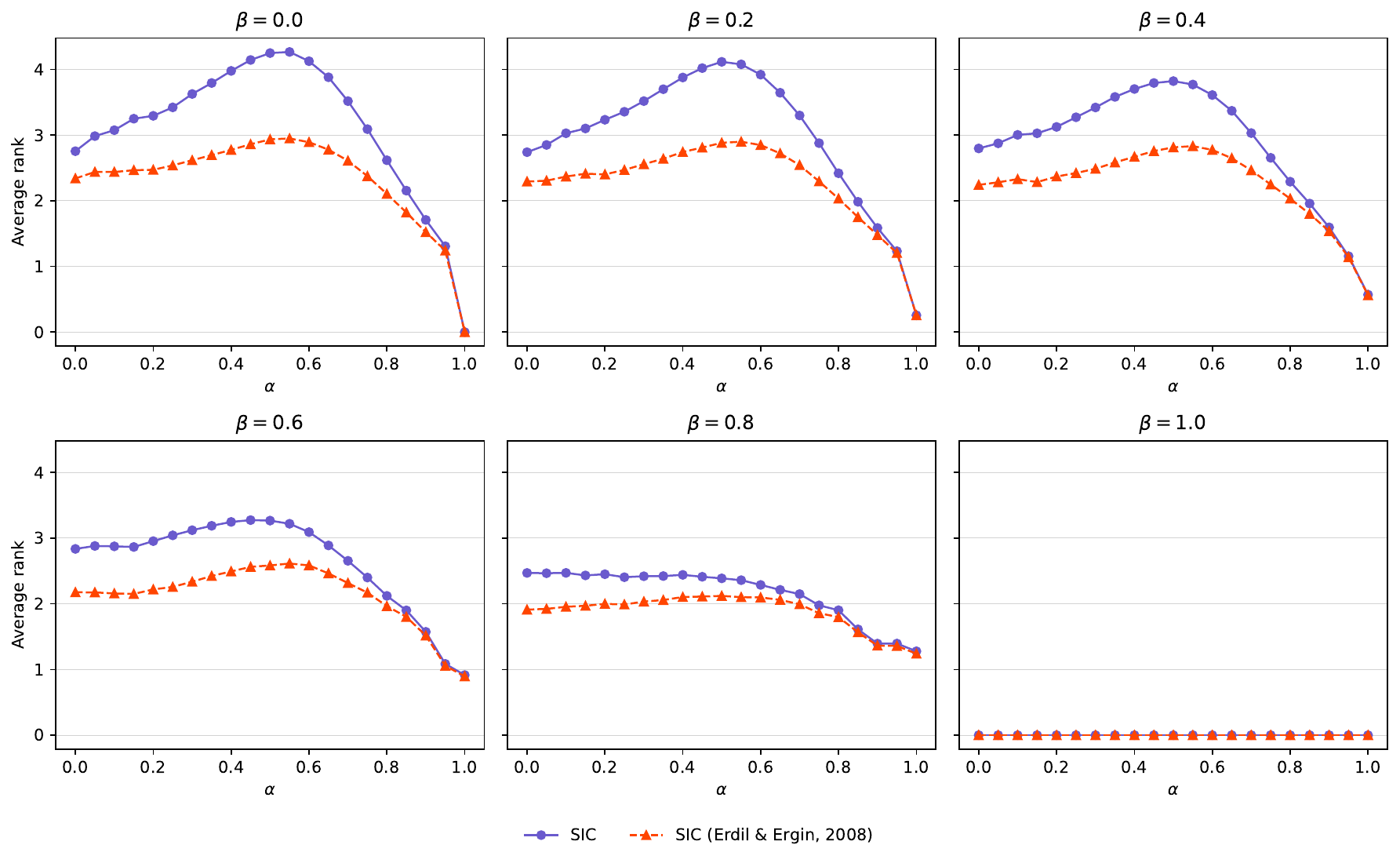}
\caption{Average improvement in rank \textit{among improving students} as a function of $\alpha$ (1,000 students and 20 schools).}
\label{fig:avg_impr_DA}

\end{figure}

\begin{figure}[!htb]
\centering
\includegraphics[width=.85\linewidth]{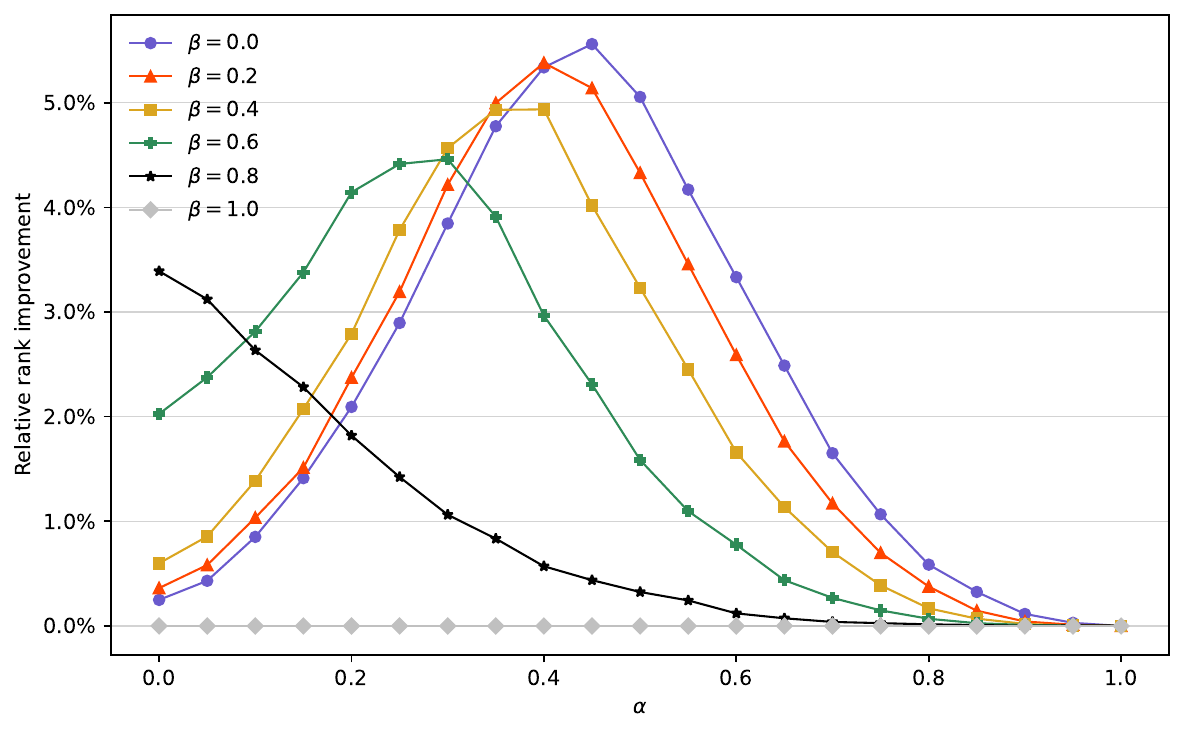}
\caption{Relative improvement in average rank of the matchings returned by corrected code, in comparison to the original code, as a function of $\alpha$.}
\label{fig:avg_rank}

\end{figure}

In only 2 of the 25,200 simulated instances, the original code resulted in an unstable matching. However, in 90.6\% of the generated instances in which the stable improvement cycles algorithm could improve upon DA, the corrected code finds a different matching than the originally implemented code. Figures \ref{fig:fraction_impr_DA} and \ref{fig:avg_impr_DA} show that while the fraction of improving students is smaller than originally reported, their average improvement in rank is larger than originally reported. Overall, the corrected code results in matchings that have a lower average rank than the original code. Figure \ref{fig:avg_rank} shows that the improvement in average rank by correcting the bug can be up to to around 5\% in comparison to the original code, depending on the values of of $\alpha$ and $\beta$.

Overall, the general insights by \cite{erdil2008s} based on the computational experiments remain valid.

\bibliographystyle{apalike}
\bibliography{bib}
\end{document}